\documentclass[lettersize,journal]{IEEEtran}
\usepackage{subcaption}
\usepackage{amsmath,amsfonts}
\usepackage{algorithmic}
\usepackage{algorithm}
\usepackage{array}
\usepackage[caption=false,font=normalsize,labelfont=sf,textfont=sf]{subfig}
\usepackage{textcomp}
\usepackage{stfloats}
\usepackage{url}
\usepackage{verbatim}
\usepackage{graphicx}
\usepackage{cite}
\usepackage{booktabs}
\usepackage{tabularx}
\usepackage{multirow}
\usepackage[table]{xcolor}
\usepackage{enumitem}
\newlist{researchquestions}{enumerate}{1}
\setlist[researchquestions]{label*=\textbf{RQ\arabic*}}
\begin{document}

\title{How Robust are LLM-Generated Library Imports? An Empirical Study using Stack Overflow}

\author{%
  Jasmine Latendresse, %
  \thanks{J. Latendresse is with Concordia University, Montreal, Canada (email: jasmine.latendresse@mail.concordia.ca).}%
  \and
  SayedHassan Khatoonabadi, %
  \thanks{S. Khatoonabadi is with Concordia University, Montreal, Canada (email: sayedhassan.khatoonabadi@concordia.ca).}%
  \and
  Emad Shihab%
  \thanks{E. Shihab is with Concordia University, Montreal, Canada (email: emad.shihab@concordia.ca).}%
}

\maketitle

\begin{abstract}

Software libraries are central to the functionality, security, and maintainability of modern code. As developers increasingly turn to Large Language Models (LLMs) to assist with programming tasks, understanding how these models recommend libraries is essential. In this paper, we conduct an empirical study of six state-of-the-art LLMs, both proprietary and open-source, by prompting them to solve real-world Python problems sourced from Stack Overflow. We analyze the types of libraries they import, the characteristics of those libraries, and the extent to which the recommendations are usable out of the box. Our results show that LLMs predominantly favour third-party libraries over standard ones, and often recommend mature, popular, and permissively licensed dependencies. However, we also identify gaps in usability: 4.6\% of the libraries could not be resolved automatically due to structural mismatches between import names and installable packages, and only two models (out of six) provided installation guidance. While the generated code is technically valid, the lack of contextual support places the burden of manually resolving dependencies on the user. Our findings offer actionable insights for both developers and researchers, and highlight opportunities to improve the reliability and usability of LLM-generated code in the context of software dependencies. 

\end{abstract}


\section{Introduction}
\IEEEPARstart{M}{odern} software development heavily relies on open source libraries that provide reusable functionalities through well-defined modules, significantly reducing development time and effort~\cite{basili1996reuse, decan2019empirical, murphy2019predicts}. While libraries can help speed up development tasks, they also introduce dependencies --- interconnections between code components --- that can lead to increased complexity and dependency management challenges~\cite{Latendresse_ASE2022, mujahid2021effective, wang2020empirical}.

One critical aspect of dependency management is library selection~\cite{mujahid2023characteristics}. Choosing the right library impacts factors like code maintainability, performance, and security. Previous studies have explored how developers select libraries, and highlighted primarily ad-hoc processes based on past experiences, expert advice, and online resources~\cite{hauge2009empirical, haenni2013categorizing}. 

Decisions around dependency adoption are influenced by factors such as functionality, community support, and maintenance compatibility~\cite{mens2024overview, vargas_selecting, mujahid2023characteristics}. Previous studies reveal that maintenance tasks like updates and vulnerability migration are often hindered by risks of breaking changes~\cite{bogart2021and, mezzetti2018type, moller2019model} and complexity from transitive dependencies~\cite{soto2021comprehensive, cao2022towards, jafari2021dependency, soto2023coverage}. Automation and tooling (e.g., Dependabot~\footnote{\url{https://github.com/dependabot}}) have been introduced to alleviate these issues, but they frequently fall short because of alert fatigue, false positives, or lack of actionable context~\cite{mohayeji2023investigating, alfadel2021use, he2023automating, pashchenko2020vuln4real, jia2021depowl}.

In parallel, the growing adoption of LLMs as programming assistants introduces new possibilities for addressing these challenges. LLMs are increasingly used to assist with code generation, and studies show their potential to enhance productivity through capabilities like code completion and search~\cite{liang2024large, ross2023programmer, heitz2024evaluation}. However, their impact on software dependencies (i.e., their ability to generate reliable library imports) remains unexplored. As these models become central in development workflows, it is critical to assess the correctness of their outputs and their implications for dependency management.

Our work aims to explore a specific aspect of LLM functionality: their ability to recommend software libraries during code generation. In this paper, we evaluate the effectiveness of six state-of-the-art LLMs by conducting an empirical study in which each model generates Python code for a set of real-world coding questions derived from Stack Overflow~\cite{stackoverflow}. This setup enables us to assess LLM behavior in a setting that closely reflects practical developer use. By analyzing the libraries recommended by these models, we aim to answer the following research questions: 

\begin{researchquestions}
    \item{} \textbf{What types of software libraries are recommended by LLMs?} Software libraries play a critical role in the security, performance, and maintainability of software projects. Thus, we examine the types of libraries recommended by LLMs and how frequently they rely on Python's standard library versus third-party. Our results suggest that LLMs tend to recommend third-party libraries more often than standard libraries. Across all models evaluated, 54\% of imports were third-party, 41\% were standard, and 5\% were unresolvable. This suggests a strong reliance on external packages rather than built-in functionality provided by Python. 

    \item{} \textbf{What are the characteristics of third-party libraries recommended by LLMs?} Third-party libraries introduce external dependencies that require separate installation and ongoing maintenance, which makes their quality and licenses critical in production contexts. To assess whether the libraries imported in LLM-generated code are correct and sustainable, we evaluate each third-party dependency along three key dimensions: popularity (e.g., GitHub stars, dependents), maintenance (e.g., age, update frequency), and licensing. Our findings show that LLMs tend to favour well-established and permissively licensed libraries that have high community adoption and low maintenance overhead. 

    \item{} \textbf{Why do some libraries recommended by LLMs not work out of the box?} Some libraries recommended by LLMs do not work out of the box, which can be disruptive to the development workflow and negatively affect trust in AI-based programming assistants. In this RQ, we investigate the root causes of these unresolved imports by analyzing cases where the recommended library could not be mapped to a known standard or third-party package. We found that most of the unresolved imports stem from structural conventions in the Python programming language, where the import name differs from the package name. We also identified one case involving a module-level import error. While the code is valid, missing installation instructions limit usability. Only a few models, such as DeepSeek V3, included installation guidance. 
\end{researchquestions}

Our findings show how effectively LLMs recommend software libraries and the factors that influence their behavior in this context during code generation. This contributes to a better understanding of model capabilities and limitations, and provides actionable insights for developers integrating LLMs into their workflows. We also highlight areas where models consistently produce reliable outputs, as well as scenarios where users should be cautious due to missing context or unresolved dependencies. Our work makes the following contributions: 

\begin{itemize}
    \item Empirical evidence on the characteristics of libraries used in LLM-generated code and the potential challenges in using these libraries out of the box.
    \item A set of actionable recommendations for developers using LLMs to streamline their workflows, including guidance for potential issues with LLM-suggested libraries.
    \item A dataset of LLM-generated Python code from six state-of-the-art LLMs that enables a direct comparison of code outputs and library selection practices, available online\footnote{\url{https://zenodo.org/records/15880965}}.
\end{itemize}

\noindent\textbf{Paper Organization.} The rest of the paper is organized as follows. Section~\ref{sec:study_design} describes our dataset and methodology. Sections~\ref{sec:rq1}, \ref{sec:rq2}, \ref{sec:rq3} present the findings of our three research questions. Section~\ref{sec:discussion} discusses the implications of our findings along with our recommendations. Section~\ref{sec:related_works} discusses related work. Section~\ref{sec:threats} outlines the threats to the validity of our study. Finally, Section \ref{sec:conclusion} concludes this paper.

\section{Study Design}
\label{sec:study_design}
\begin{figure*}[t]
    \centering
    \includegraphics[scale=0.45]{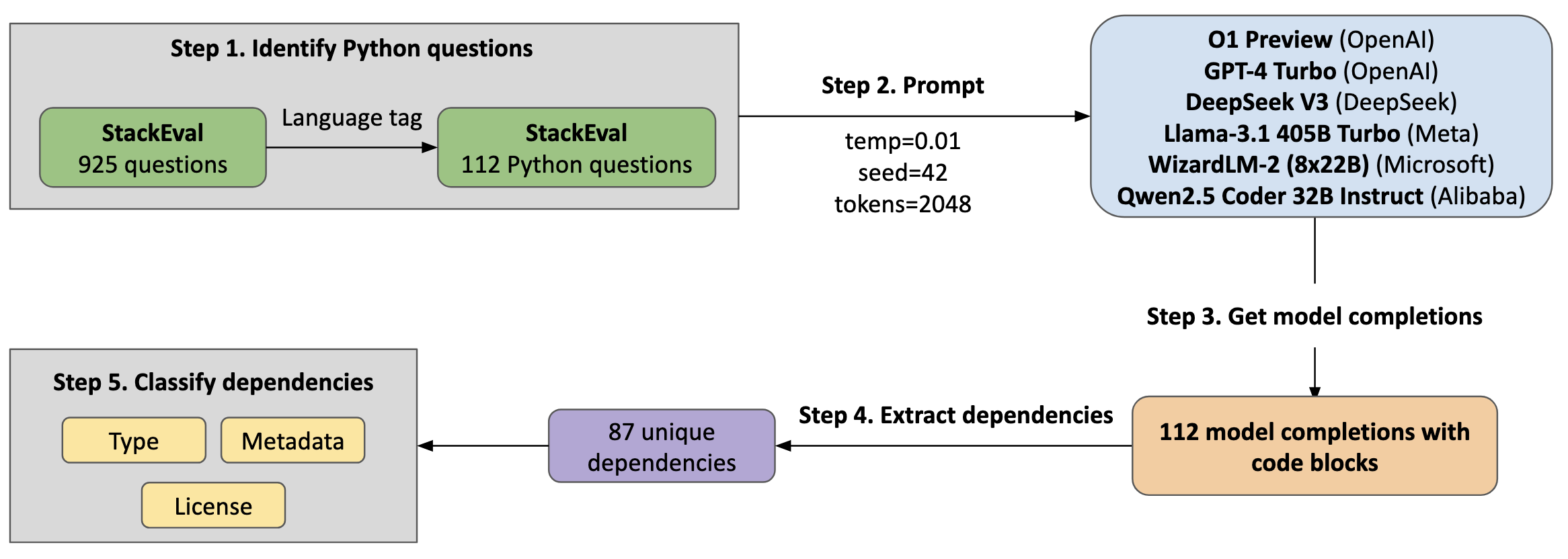}
    \caption{An overview of our approach for filtering and curating our dataset.}
    \label{fig:approach}
\end{figure*}

In this section, we describe the steps to curate the dataset used in our study and the experimental setup. 

\subsection{Dataset}
\label{sec:dataset}
Figure~\ref{fig:approach} presents an overview of our approach for curating the dataset used in this study and analyzing the dependencies generated by LLMs. We use the StackEval benchmark suite curated by Shah et al.~\cite{shah2024stackeval}, which contains question-answer pairs sourced from Stack Overflow. The benchmark is designed to evaluate LLM performance on coding tasks and includes 925 questions sampled between January 2018 and September 2023. Each question includes a title, body, at least one upvote, and an accepted answer that also received one upvote.

We selected this dataset because Stack Overflow questions reflect real-world programming scenarios where developers seek help with writing, fixing, or improving code. These questions often involve choosing or using software libraries, whether explicitly (e.g., Difference between numpy power and ** for certain values) or implicitly (i.e., through the context of a coding task). Thus, using Stack Overflow questions allows us to assess how LLMs respond in practical contexts, where library recommendation naturally occurs as part of solving coding problems. 

We filtered the dataset to retain Python-related questions using the language tags. Python was chosen due to its popularity, rich ecosystem of third-party libraries, and the strong performance of LLMs in generating and understanding Python code~\cite{saabith2020popular, srinath2017python, adamson2023assessing, ellis2024chatgpt}. This filtering resulted in 112 Python programming questions.

\subsection{Experiment Setup}
\label{sec:exp_setup}

We evaluate six state-of-the-art LLMs, including two proprietary models and four open-source models. The proprietary models are O1 Preview~\cite{zhong2024evaluation} and GPT-4 Turbo~\cite{achiam2023gpt}, both developed by OpenAI. The open-source models include DeepSeek V3~\cite{liu2024deepseek} from DeepSeek AI, Llama 3.1 405B Turbo~\cite{grattafiori2024llama} from Meta, WizardLM-2 8x22B~\cite{xu2023wizardlm} from Microsoft, and Qwen 2.5 Coder 32B Instruct~\cite{hui2024qwen2} from Alibaba. Table~\ref{tab:model_overview} provides an overview of their key characteristics and capabilities.

\begin{table*}[t]
\centering
\caption{Overview of Evaluated LLMs}
\label{tab:model_overview}
\begin{tabular}{l l l l l l}
\toprule
\textbf{Model} & \textbf{Provider} & \textbf{Nb. Parameters} & \textbf{Publication Year} & \textbf{Training Cutoff} & \textbf{Capabilities} \\
\midrule
DeepSeek V3     & DeepSeek AI & 685B      & December 2024 & July 2024     & Mixture-of-experts, general-purpose, code, reasoning. \\
GPT-4 Turbo     & OpenAI      & $\sim$1.8T & April 2024    & December 2023 & Multimodal, general purpose, code, reasoning. \\
Llama 3.1       & Meta        & 405B      & July 2024     & December 2023 & Instruction-tuned, general purpose, text, code. \\
O1 Preview      & OpenAI      & $\sim$175B & September 2024 & October 2023  & Reasoning-optimized. \\
Qwen2.5         & Alibaba     & 32B       & September 2024 & March 2024    & Instruction-tuned, strong support for code. \\
WizardLM-2      & Microsoft   & 141B      & April 2024    & October 2023  & Mixture-of-experts, instruction-tuned, code. \\
\bottomrule
\end{tabular}
\end{table*}

To collect model completions, we follow the approach of Shah et al.~\cite{shah2024stackeval}. Each model is prompted with the selected 112 Python questions from StackEval using only the question title and body, without additional instructions or examples. This setup reflects realistic usage, where developers typically pose natural language questions without carefully crafted prompts~\cite{chouchen2024software}, and allows us to evaluate model behavior in conditions that are minimally guided. The parameter configuration has a near-deterministic temperature (0.01), fixed seed (42), and outputs limited to 2,048 tokens. These settings follow the configuration used in the original StackEval benchmark~\cite{shah2024stackeval}, and are chosen to minimize the randomness of the output and ensure reproducibility. 

After generating responses, we extract code blocks using a regular expression. We then identify software libraries by locating import statements within the code and extracting the parent library name (e.g., from \texttt{from X import Y}, we extract \texttt{X}, from \texttt{import X}, we extract \texttt{X}, from \texttt{import X.Y}, we extract \texttt{X}). This allows us to focus on high-level usage patterns of libraries. This process results in 87 unique libraries extracted from the 112 completions, all models aggregated. In our analysis, we consider each library as an individual case regardless of how many times it appears across questions or models. For per-model analysis, libraries may repeat across different questions. For aggregated model analysis, we focus on the set of unique libraries recommended across all completions. The last step involves classifying the libraries by type, and collecting their metadata and license information. We provide a detailed description of this classification process in RQ1.

\section{RQ1: What types of software libraries are recommended by LLMs?}
\label{sec:rq1}

\noindent\textbf{Motivation.} As developers increasingly turn to AI-based programming assistants to streamline their development workflows, the choice of libraries included in AI-generated code becomes critical. These libraries directly impact the resulting code's performance, security, and functionality~\cite{Latendresse_ASE2022, jafari_update, cox2019surviving}. This research question aims to examine the types of libraries included in LLM-generated code and how frequently models rely on Python's built-in standard libraries versus third-party or unidentifiable libraries. Understanding this helps evaluate the reliability and integration complexity of LLM-generated library imports in real-world development settings.

\noindent\textbf{Approach.} We classify each imported library as one of the following: (1) \textit{standard}, meaning it is bundled with the default Python installation, (2) \textit{third-party}, meaning it is installable via the Python Package Index (PyPI), or (3) \textit{unknown}, meaning it cannot be mapped to a known standard or third-party library. To perform this classification, we use the \texttt{stdlib-list} package to identify standard libraries (for Python 3) and query the PyPI API for the rest. Libraries that are not found in either are labeled as \textit{unknown}.

\noindent\textbf{Results.} Table~\ref{tab:deps_per_model_eval} presents the distribution of dependency types. The first row shows the results for all combined models. We identified a total of 87 distinct dependencies of which 47 (54.0\%) were \textit{third-party} libraries, and 36 (41.4\%) were \textit{standard} libraries. This suggests that models are more likely to generate code that imports external packages rather than relying on built-in Python functionality. Moreover, four (4.6\%) of the libraries were classified as \textit{unknown}, indicating these libraries are neither identified as \textit{standard} nor \textit{third-party}. We further investigate this category and its implications in RQ2.

\begin{table}[h]
    \centering
    \caption{Classification of Imported Libraries in LLM-generated Code}
    \label{tab:deps_per_model_eval}
    \begin{tabular}{lcccc}
        \toprule
        \textbf{Model} & \textbf{Std.} & \textbf{3rd-Party} & \textbf{Unknown} & \textbf{Total} \\
        \midrule
        All models & 36 (41.4\%) & 47 (54\%) & 4 (4.6\%) & 87 \\
        \midrule
        DeepSeek V3       & 24 (42.1\%) & 30 (52.6\%) & 3 (5.3\%) & 57 \\
        GPT-4 Turbo            & 26 (40.6\%) & 35 (54.7\%) & 3 (4.7\%) & 64 \\
        Llama 3.1         & 22 (40.7\%) & 28 (51.9\%) & 4 (7.4\%) & 54 \\
        O1 Preview        & 18 (39.1\%) & 26 (56.5\%) & 2 (4.3\%) & 46 \\
        Qwen2.5           & 21 (36.2\%) & 34 (58.6\%) & 3 (5.2\%) & 58 \\
        WizardLM-2        & 19 (38.8\%) & 28 (57.1\%) & 2 (4.1\%) & 49 \\
        \bottomrule
    \end{tabular}
\end{table}

Looking at the results per model, we observe a consistent pattern across all six LLMs: third-party libraries account for the largest proportion of imports. GPT-4 Turbo recommends the most unique libraries overall (64), with \textit{standard} libraries accounting for 40.6\%, third-party accounting for 54.7\%, and three instances of unknown imports. Llama 3.1 has the highest proportion of \textit{unknown} libraries (5), though this is only marginally higher than the other models, which range between 2 and  4. This small variation is not substantial enough to indicate systematic differences between models. Nonetheless, its proportions of \textit{standard} and \textit{third-party} imports remain comparable to other models. O1 Preview has the lowest total number of distinct dependencies (46), but maintains a similar overall trend, with a majority of third-party recommendations (56.5\%). Qwen2.5 has the highest proportion of \textit{third-party} dependencies (58.6\%), which could have implications for integration complexity and maintenance. WizardLM-2 has the smallest proportion of unknowns (4.1\%), but also shows a high third-party usage rate (57.1\%). Overall, we can see that all models tend to use \textit{third-party} libraries more often than \textit{standard} with varying degrees (e.g., Qwen2.5, WizardLM-2).

\begin{center}\setlength{\fboxsep}{10pt}
\setlength{\fboxrule}{1pt}
\fcolorbox{gray!60}{gray!20}{%
    \parbox{0.9\columnwidth}{%
        Our analysis shows that all six LLMs generate code that predominantly relies on third-party libraries, ranging between 51.9\% and 58.6\% of imports across models. Out of 87 distinct libraries identified across all completions, 54.0\% were third-party, 41.4\% were standard, and 4.6\% were unknown. GPT-4 generated the broadest set of libraries, while Llama 3.1 has the highest proportion of unknowns. 
    }  
    }
\end{center}

\section{RQ2: What are the characteristics of third-party libraries recommended by LLMs?}
\label{sec:rq2}

\noindent\textbf{Motivation.} Unlike standard libraries, third-party libraries introduce external dependencies that must be installed and maintained separately. Their characteristics, such as popularity, maintenance activity, and licensing, are known to significantly affect the long-term stability and legal compliance of software projects~\cite{mens2024overview}. Thus, in this research question, we aim to evaluate whether LLMs tend to recommend third-party libraries that are mature, reliable, and safe for production.

\noindent\textbf{Approach.} We analyze each third-party library along three key dimensions based on prior work~\cite{jafari_update, vargas_selecting}: popularity, maintenance, and licensing. For popularity, we examine the number of GitHub forks, stars, and dependent projects. For maintenance, we look at the number of dependencies, age (calculated from the library's first release date to the metadata collection date (February 2024)), version release frequency, and SourceRank score (provided by libraries.io). Finally, we assess licensing information by assigning SPDX license identifiers and grouping them into categories such as permissive, copyleft, and weak copyleft. 

\noindent\textbf{Results.} Table~\ref{tab:dependency_medians_all_and_per_model} presents the median values of key dependency characteristics for each model as well at the aggregated results. Aggregated values for the mean, standard deviation, minimum, and maximum are available in the appendix. Below, we discuss the results in more detail. 

\textbf{Popularity.} The popularity metrics (forks, stars, and dependents) show how widely adopted and trusted a library is within the Python ecosystem. They reflect developer interest (e.g., stars), collaborative engagement (e.g., forks), and real-world usage (e.g., dependents). The aggregated medians across models show that LLMs generally recommend libraries that are well-established and commonly used, with a median number of forks of 1,473, 6,962 stars, and 1,140 dependents. 

At the per-model level, we observed that the median number of stars ranges from 5,471 to 8,528. This indicates that all models recommend libraries with a notable level of visibility and level of adoption. As shown in Table~\ref{tab:dependency_medians_all_and_per_model}, some models fall toward the high end of this range (e.g., Qwen2.5 and Llama 3.1), and some fall toward the lower end (e.g., WizardLM-2 and DeepSeek V3), suggesting that they might recommend more specialized or context-specific libraries. Other models have popularity metrics that are close to the overall medians (e.g., GPT-4 Turbo on all popularity metrics and O1 Preview on forks and dependents), which suggests that their recommendations are still broadly visible and adopted, but not skewed towards highly popular or niche options. 

\textbf{Maintenance.} In terms of maintenance, the results show that most models recommend libraries that are mature and self-contained. The median number of dependencies across all models is low (1.0) and never exceeds 3.0 at the per-model level. This indicates that the recommended libraries are lightweight in terms of transitive dependencies, reducing integration risk and potential for version conflicts in downstream projects~\cite{mens2024overview}. 

The SourceRank metric, which reflects a library's overall quality and development activity, is relatively consistent across models, ranging from 22 to 24, with an overall median of 23 (out of a maximum of 30). This indicates that LLMs tend to select libraries that are well-maintained and well-regarded in the community. The age of recommended libraries is consistently high, with median values across models above 7 years, which suggests LLMs tend to recommend stable, long-standing libraries over newer ones. From a developer's perspective, this behaviour may promote trust and ensure compatibility, especially in production environments where stability and maturity are important.

Version release frequency remains below one release per month for all models, with medians ranging from 0.58 to 0.77 (\~17 to 23 days). These values imply that most libraries receive regular but not overly frequent updates, which is often preferred by developers who want to avoid breaking changes, frequent manual interventions, or unexpected regressions~\cite{pashchenko2020qualitative}. The observed range for version frequency does not appear to correlate directly with the training cutoff dates. For instance, O1 Preview (cutoff: October 2023) and Qwen2.5 (cutoff: March 2024) both recommend libraries with identical median version frequency per month (0.77). This suggests that other factors like training data composition, domain coverage, or fine-tuning strategies may have a greater influence on the kinds of libraries selected by each model. 

\begin{table*}[t]
\centering
\caption{Median values of recommended library characteristics.}
\label{tab:dependency_medians_all_and_per_model}
\begin{tabular}{llccccccc}
\toprule
\textbf{Category} & \textbf{Metric} 
& \textbf{All Models} 
& \textbf{DeepSeek V3} 
& \textbf{GPT-4 Turbo} 
& \textbf{Llama 3.1} 
& \textbf{O1 Preview} 
& \textbf{Qwen2.5} 
& \textbf{WizardLM-2} \\
\midrule
\multirow{3}{*}{Popularity} 
& Forks      & 1,473 & 1,292 & 1,473 & 1,682 & 1,682 & 1,430 & 1,292 \\
& Stars      & 6,962 & 5,607 & 6,962 & 8,528 & 8,073 & 8,073 & 5,471 \\
& Dependents & 1,140 & 1,068 & 1,114 & 1,068 & 1,287 & 2,080 & 1,287 \\
\midrule
\multirow{4}{*}{Maintenance}
& Dependencies       & 1.0 & 0.5 & 2.0 & 2.5 & 0.0 & 1.5 & 1.5 \\
& SourceRank         & 23 & 22.5 & 23 & 22 & 22.5 & 24 & 22 \\
& Age (months)       & 95 & 94.9 & 97.7 & 92.9 & 87.3 & 99.5 & 104.1 \\
& Version Frequency  & 0.64 & 0.62 & 0.58 & 0.62 & 0.77 & 0.77 & 0.77 \\
\bottomrule
\end{tabular}
\end{table*}

\begin{figure*}[]
    \centering
    \includegraphics[scale=0.6]{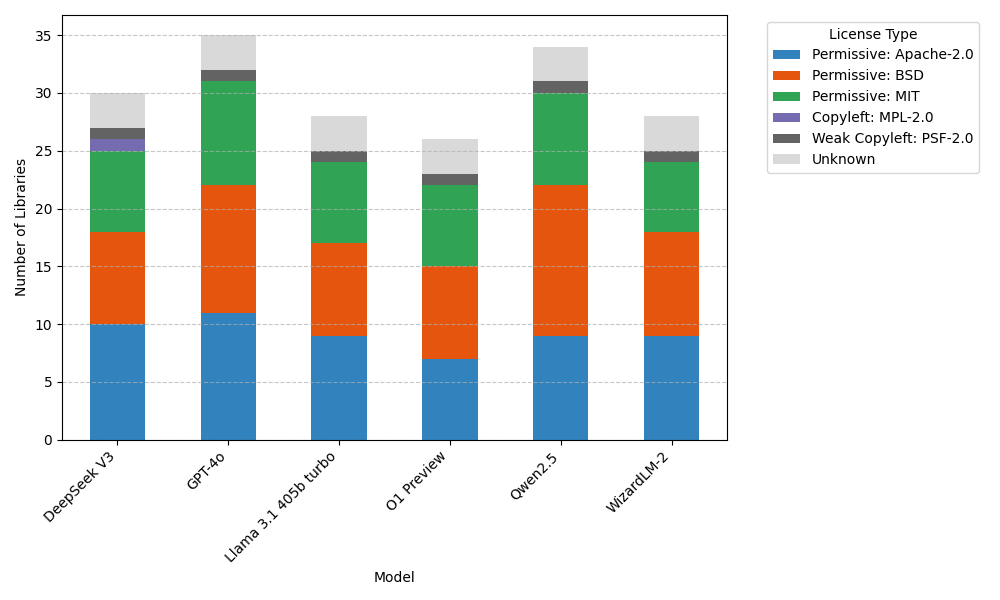}
    \caption{License Types in Libraries Recommended by Each Model}
    \label{fig:license_type}
\end{figure*}

\textbf{License.} Figure~\ref{fig:license_type} shows a breakdown of the licenses associated with the third-party libraries recommended by each model. The majority of the libraries are distributed under permissive licenses. On a total of 181 license instances across all models, 164 (91\%) fall into the permissive category, including licenses such as Apache-2.0, BSD, and MIT. Among these, Apache-2.0 and BSD are the most common. It is worth noting that we consolidated subtypes of BSD licenses (e.g., BSD-2-clause and BSD-3-clause) into a single BSD category. DeepSeek V3 is the only model that recommended a copyleft-licensed library (\texttt{certifi}), which requires derivative works to also remain open source. Each model recommended a library under the weak-copyleft PSF-2.0 license (\texttt{matplotlib}), which offers more flexibility than copyleft by allowing derivative works to be closed. 

However, each model consistently recommends the same three libraries for which the license is not explicitly declared on PyPI (\textit{unknown}): \texttt{myapp}, \texttt{sklearn}, and \texttt{imblearn}. For \texttt{myapp}, the documentation indicates that it is a test library~\footnote{\url{https://pypi.org/project/myapp/}}, which could explain why no license is explicitly declared. We observed that the models use this name as a placeholder instead of referring to the actual library. We discuss this occurrence in more detail in RQ2. 

In the case of \texttt{sklearn}, the corresponding PyPI page indicates that the library is deprecated~\footnote{\url{https://pypi.org/project/sklearn/}}, which led us to classify it as having no declared license. However, we found that \texttt{sklearn} is an import alias for the actively maintained \texttt{scikit-learn} library. This means that while developers install the library using the name \texttt{scikit-learn}, it is imported in Python as \texttt{sklearn}. As a result, querying \texttt{sklearn} directly on PyPI leads to a deprecated library. In reality, \texttt{scikit-learn} is distributed under a permissive type of license (BSD). We discuss the implications of such aliasing further in RQ2. Finally, for \texttt{imblearn}, the PyPI documentation recommends using \texttt{imbalanced-learn} instead~\footnote{\url{https://pypi.org/project/imblearn/}}, yet no explicit deprecation notice is shown and unlike \texttt{scikit-learn}, we found no evidence of aliasing for this case in our dataset. Thus, the results show that LLMs recommend libraries that are widely permissive, which is ideal in policy-sensitive production settings, but there are a few cases where the license can be more restrictive (e.g., copyleft) or not explicitly declared, which can lead to potential legal challenges. Practitioners should actively verify the license of the libraries recommended by LLMs they integrate into their codebases.  

\begin{center}\setlength{\fboxsep}{10pt}
\setlength{\fboxrule}{1pt}
\fcolorbox{gray!60}{gray!20}{%
    \parbox{0.9\columnwidth}{%
        Our findings show that LLMs predominantly generate dependencies on third-party libraries that are popular, well-established, and permissively licensed. Across all models, the recommended libraries are stable (median version frequency of 0.64), have minimal transitive dependencies (median of 1 dependency), and are well-regarded by the community (median SourceRank of 23).
    }  
    }
\end{center}

\section{RQ3: Why do some libraries recommended by large language models not work out of the box?}
\label{sec:rq3}

\noindent\textbf{Motivation.} In RQ1, we observed that 4.6\% of libraries in LLM-generated code do not function out of the box, meaning that they are neither available on PyPI (i.e., \textit{third-party} library) nor included in the standard Python distribution (i.e., \textit{standard} library). This can lead to confusion, hinder software development workflows, and negatively affect trust in LLM-generated code~\cite{adamson2023assessing}. Thus, in this RQ, we examine the underlying reasons why such libraries are included in LLM-generated code. Understanding this is important for improving the reliability of LLMs in programming tasks and their practical utility in real-world settings. RQ2 therefore aims to identify and characterize the common sources of non-functional library recommendations in LLM outputs. 

\noindent\textbf{Approach.} To understand why some libraries recommended by LLMs do not work out of the box, we analyzed each case identified as \textit{unknown} in RQ1 (i.e., neither \textit{third-party} nor \textit{standard}). Specifically, we qualitatively examined the context in which each library was recommended by analyzing the associated Stack Overflow question used in the prompt, including its body, code snippets, and accepted answer. This process allowed us to infer whether the recommendation stemmed from a misunderstanding of naming conventions, structural issues in the import statement, or other contextual factors. Based on this, we developed a set of mutually exclusive categories that characterize the underlying reasons why certain libraries in LLM-generated code fail to resolve correctly. Since this categorization relied on concrete, factual properties, the first author performed the initial labeling, and the results were reviewed and validated by the second author. This led to the identification of two primary categories: 

\begin{itemize}
    \item \textbf{Alias}: The extracted library name corresponds to an alias used in Python import statements, rather than the actual name of the library used for installation. For example, \texttt{sklearn} is used as an alias for the \texttt{scikit-learn} library. While the model's suggestion is functionally correct, alias-based imports may lead to errors during dependency resolution.
    \item \textbf{Module}: The extracted name refers to an internal module or subcomponent within a third-party library. These cases occur when the model suggests imports from a specific file using non-standard or path-based import patterns. These cases also happen with implicit imports, which is when a module is imported directly without referencing the parent library. Since the name used in the import does not match the actual installable library, it can lead to a failure in dependency resolution if the parent library is not installed. 
\end{itemize}

\noindent\textbf{Results.} Table~\ref{tab:alias_module_cases} shows the \textit{unknown} libraries identified in RQ1, categorized into the two underlying causes discussed above: \textit{alias} and \textit{module}. The \textit{alias} category accounts for most of the \textit{unknown} cases across all six models, with a total of 24 occurrences (3 unique). The most frequent alias is \texttt{cv2} used for the \texttt{opencv-python} library, which appears in all models. Similarly, \texttt{dateutil} is used as an alias for \texttt{python-dateutil}, also appearing in all models. The last alias, \texttt{yaml}, used for the library \texttt{PyYAML}, appears in half of the models (GPT-4 Turbo, Llama 3.1, and Qwen2.5). 

These imports are functionally correct, but can lead to confusion since the alias used in the import statement does not match the actual installable library name. To further investigate the practical consequences of this, we manually reviewed the responses generated by the models for each of the 24 alias cases. We found that only 5 of these completions included an installation step (e.g., \texttt{pip install opencv-python}), which occurred in outputs from DeepSeek V3, GPT-4 Turbo, and Qwen2.5. This indicates that models may often assume implicit knowledge about the correct library to install and places the burden on developers to manually infer the library before they execute the code. This is especially important in educational contexts, where users may lack the knowledge to resolve missing dependencies. 

In contrast to \textit{alias} cases, we identified a single instance categorized as \textit{module}. This occurred in a response generated by Llama 3.1, where the model produced the statement \texttt{from client import EmailageClient}, which resulted in the extraction of \texttt{client} as the top-level dependency. However, \texttt{client} is not a standalone installable library; it is a submodule nested within the \texttt{emailage} library. Thus, the import statement is not incorrect in itself, but attempting to install \texttt{client} on its own would fail. This is a challenge that occurs with implicit imports: the module is imported without an explicit reference to its parent library. In this particular case, Llama 3.1 did not provide additional installation steps in the completion, leaving it to the user to manually infer the library to install. It could be argued that the user might possess the necessary knowledge given that they formulated the original question, but this assumption is problematic. Similar to the alias cases, it places the burden of dependency resolution on the user and undermines the practical reliability of LLMs in real-world settings. 

\textbf{Per-model analysis.} Although all models produce some \textit{unknown} libraries, their behavior varies slightly. For instance, GPT-4 Turbo, Qwen2.5, and Llama 3.1 tend to rely more heavily on alias imports. However, these aliases (e.g., \texttt{cv2}, \texttt{dateutil}, \texttt{yaml}) are widely adopted and well-established libraries. Interestingly, only these three models used the \texttt{yaml} alias, despite all models receiving the same input, which suggests that they may have a more accurate understanding of the \texttt{PyYAML} library to recommend it for the same use case. In contrast, WizardLM-2 and O1 Preview produced the fewest \textit{unknowns}, which could suggest they are more conservative about importing libraries. This is reflected in RQ1, where these models produced the fewest unique imports overall. 

DeepSeek V3, which produced a moderate number of aliases (4), was one of the few models that included installation guidance (with GPT-4 Turbo and Qwen2.5), indicating that the model can both produce correct imports and correctly map aliases to their installable libraries. This may indicate that models which include installation guidance in their completion anticipate potential friction during code execution and proactively address it. While this does not guarantee robustness, it suggests a model-level difference in how much contextual support is embedded model's output, a feature that enhances usability and trust. 

Our results show that all models are capable of generating technically correct code, but some models may place a greater burden on the user to interpret and resolve dependencies, especially when explicit installation instructions are omitted from the output. 

\begin{table*}
\centering
\caption{Occurrences of alias and module cases per model.}
\label{tab:alias_module_cases}
\begin{tabular}{lclcl}
\toprule
\textbf{Model} & \textbf{\# Alias Occ.} & \textbf{Alias instances} & \textbf{\# Module Occ.} & \textbf{Module instances} \\
\midrule
DeepSeek V3 & 4 & \texttt{cv2} (1), \texttt{dateutil} (3) & 0 & -- \\
GPT-4 Turbo & 5 & \texttt{cv2} (2), \texttt{dateutil} (2), \texttt{yaml} (1) & 0 & -- \\
Llama 3.1 & 5 & \texttt{cv2} (1), \texttt{dateutil} (2), \texttt{yaml} (1) & 1 & \texttt{client} (1) \\
O1 Preview & 3 & \texttt{cv2} (1), \texttt{dateutil} (2) & 0 & -- \\
Qwen2.5 & 5 & \texttt{cv2} (1), \texttt{dateutil} (3), \texttt{yaml} (1) & 0 & -- \\
WizardLM-2 & 2 & \texttt{cv2} (1), \texttt{dateutil} (1) & 0 & -- \\
\midrule
\textbf{Total} & 24 & & 1 & \\
\bottomrule
\end{tabular}
\end{table*}

\begin{center}\setlength{\fboxsep}{10pt}
\setlength{\fboxrule}{1pt}
\fcolorbox{gray!60}{gray!20}{%
    \parbox{0.9\columnwidth}{%
        Our findings show that the majority of unresolved library references stem from aliasing, where the import statement refers to a library name that differs from its installable package name. We also identifid one case involving an implicit module import, in which a submodule of a library is referenced without its parent library. We conclude that while the code produced is syntactically correct, it lacks sufficient context to ensure seamless execution and places a burden on users to resolve dependencies manually. Only a few models, such as DeepSeek V3, offered installation guidance, highlighting key differences in practical usability across models. 
    }  
    }
\end{center}

\section{Discussion}
\label{sec:discussion}
In this section, we discuss the implications of our work and propose recommendations for practitioners and researchers in the context of LLMs and dependencies. 

\subsection{How do LLMs perform on unseen questions?}
\label{sec:unseen}

A central threat to validity in studies using LLM-generated code is the possibility of data leakage (i.e., models having been exposed to the same or similar questions during training). To address this, we repeated our analysis on StackUnseen, a separate dataset specifically curated to reduce this risk~\cite{shah2024stackeval}. This dataset complements StackEval by focusing on more recent Stack Overflow questions (as of October 2024), which are less likely to have been present in the models' training corpora. From the 494 questions in StackUnseen, we extracted 71 Python-related questions and applied the same methodology as used in the StackEval analysis, which yielded a set of 66 unique libraries. 

The goal of this analysis is to determine whether model behavior changes when faced with unseen questions, specifically regarding the characteristics of the libraries they recommend. A significant shift in the characteristics of the recommended libraries could indicate a reliance on previously seen data and would raise concerns about the generalizability of our results. 

To evaluate this, we conducted a Mann-Whitney U test, a non-parametric statistical test used to compare distributions without assuming normality. We compared the distributions of library characteristics between StackEval and StackUnseen at the aggregated level (i.e., considering only unique third-party libraries for all models). The characteristics assessed are the number of stars, forks, dependents, dependencies, SourceRank, age (in months), and version frequency.

\begin{table}[h]
\centering
\caption{Mann-Whitney U Test Results Comparing StackEval and StackUnseen Library Characteristics}
\label{tab:stat-test}
\begin{tabular}{lrrrr}
\toprule
\textbf{Metric} & \textbf{Median (Eval)} & \textbf{Median (Unseen)} & \textbf{U} & \textbf{p-value} \\
\midrule
Stars & 6,962 & 10,400 & 778.5 & 0.41 \\
Forks & 1,473 & 2,032 & 829 & 0.72 \\
Dependents & 1,140 & 1,520 & 748.5 & 0.28 \\
Dependencies & 1 & 2 & 755.5 & 0.28 \\
SourceRank & 23 & 24 & 773 & 0.39 \\
Age (months) & 95 & 102 & 816.5 & 0.64 \\
Version Freq. & 0.64 & 0.89 & 713.5 & 0.16 \\
\bottomrule
\end{tabular}
\end{table}

Table~\ref{tab:stat-test} shows the results of the Mann-Whitney U test for each characteristic, along with the median values observed in StackEval and StackUnseen, all models aggregated. Across all characteristics, the results show no statistically significant differences (all $p$-values $> 0.05$). For example, the median number of stars was 3,308 for StackEval and 4,206 for StackUnseen, with a p-value of 0.41. Similarly, the median number of dependencies was 2 for StackEval and 3 for StackUnseen (p = 0.28). These results show that while there are slight shifts in central tendency, they are not statistically significant. Thus, the models consistently recommend libraries of similar maintainability and popularity, regardless of whether the input questions are historical or recent, which provides additional confidence that our original findings based on StackEval are robust and not substantially affected by potential data leakage. 

In addition to evaluating the characteristics of recommended libraries (RQ1), we examined the subset of libraries that were identified as \textit{unknown} (i.e., not clearly identifiable as \textit{standard} or \textit{third-party}). These represent edge cases where model recommendations may not work out of the box and require manual resolution from the user. As seen in RQ2, in StackEval, we identified 24 such cases (4.3\% of the recommendations), primarily due to aliasing and module-level imports. 

In contrast, StackUnseen yielded only 4 \textit{unknown} cases (1.2\%). Aliasing accounts for 50\% of these cases (\texttt{browsermobproxy}, \texttt{seleniumrequests}), while placeholder usage accounts for the other 50\% (\texttt{some\_module}). In this context, a placeholder refers to a stand-in or generic example used by the model that is meant to be replaced by the user with a real library. This reduction in \texttt{unknown} cases may be due to models defaulting to more popular or generic imports when faced with truly unseen data. It could also be partially influenced by the smaller sample size in StackUnseen, which reduced the likelihood of observing edge cases. Therefore, it is not possible to draw firm conclusions about the influence of unseen questions on this outcome. 

Our findings have practical implications for developers relying on LLM-based code generation tools. First, the consistency in library characteristics between StackEval and StackUnseen shows models exhibit relatively robust behavior in their library recommendations, even when answering questions unlikely to have been seen during training. This reinforces the reliability of our findings and supports the use of StackEval as a benchmark for future evaluations. Second, although rare, the presence of \textit{unknown} cases, even in unseen data, highlights the need for caution when integrating LLM-generated code into software projects. Developers should be aware that LLMs may suggest libraries that do not install cleanly or require specific domain knowledge to resolve. Incorporating mechanisms that prompt LLMs to clarify or justify their library usage could improve usability in real-world development workflows. 

\subsection{Practical Considerations for LLMs in the Context of Dependencies}
\label{sec:considerations}

Our study provides several insights for both practitioners and researchers who integrate LLMs into software development workflows. Below, we discuss insights relating to legal compliance, reliability, dependency management, and the overall robustness of LLM-generated library imports. 

\noindent\textbf{Licensing Awareness.} Developers should be mindful of the licensing implications of LLM-recommended libraries. Although our study shows that the vast majority of libraries suggested by the models have a permissive license (e.g., MIT, BSD, Apache-2.0), we observed a recommendation involving a copyleft-licensed library (MPL-2.0). While such occurrences are rare according to our results, they still require manual verification by the developer to ensure compliance with licensing terms, since LLMs typically do not disclose the license of the libraries they recommend. This is especially important in proprietary or commercial contexts, as failure to comply can result in legal consequences~\cite{wolter2023open}. Future work could expand this investigation beyond the Python ecosystem (e.g., JavaScript, Java, Rust) to assess whether similar licensing patterns or risks emerge across languages. 

\noindent\textbf{Users as Active Evaluators.} The results of RQ2 suggest that the failure of certain library imports to work out of the box is not tied to the models' performance, but rather is a side effect of the PyPI ecosystem itself. Thus, the successful execution of the import statement often depends on implicit conventions or contextual knowledge (e.g., aliasing, module-level imports). However, the burden remains on the user to resolve these cases. Our manual inspection revealed that installation instructions were rarely included in the model completions (only 5 out of 24 alias cases offered guidance), which leaves developers to deduce the appropriate library independently. This is particularly problematic in educational contexts, where users may not have the domain expertise to identify and correct such issues. This further highlights the importance of treating LLM outputs as a starting point rather than a definitive solution and the need for support mechanisms (e.g., tooling or documentation cues) to help developers validate and act on LLM-generated recommendations. 

\noindent\textbf{Impact on Dependency Management.} Beyond correctness, LLM-generated library imports can influence the complexity and maintainability of the codebase. Our analysis showed that LLMs tend to use third-party libraries more frequently than standard libraries, which introduces additional complexity and risks into software projects~\cite{wang2020empirical, raemaekers2011exploring}. Each third-party library introduces its own set of dependencies, the majority of which are transitive dependencies that are automatically included but not explicitly visible to the developer~\cite{Latendresse_ASE2022}. While the libraries recommended by the models were generally mature, popular, and well-maintained, the increased reliance on external packages can contribute to long-term maintenance challenges~\cite{pashchenko2020qualitative}. These include dependency hell~\cite{mens2024overview}, where version conflicts and deep dependency chains make upgrades difficult, and update churn~\cite{jafari2021dependency}, where libraries change frequently and may introduce breaking changes or require adaptation. Currently, there are no mechanisms to highlight these risks at generation time, and developers may unknowingly accumulate fragile or hard-to-maintain dependencies. Thus, there is a need for LLM-integrated tools that incorporate maintenance indicators (e.g., update frequency, dependency count, or semantic versioning signals) to help developers make informed decisions about adopting suggested libraries. 

\noindent\textbf{Security Implications.} While our study reports no hallucinated libraries, such libraries still pose a non-trivial risk that is worth discussing in the context of this paper. Recent work has shown that hallucinations can be exploited by malicious actors through dependency confusion attacks~\cite{lanyado-2023}. In such cases, a model may recommend a non-existent package, which a malicious actor could then register under the same name in a public repository. If a user installs the package without verifying its source, they risk introducing malicious code into their environment. In RQ2, we identified multiple aliased libraries, and Section~\ref{sec:unseen} revealed a placeholder import (e.g., \texttt{some\_module}), neither of which map to installable packages. While these cases are contextually correct, they still create an opportunity for exploitation if users do not recognize them as non-functional.

\textbf{For practitioners}, this reinforces the need to serve as active evaluators when interacting with LLM-generated code, especially regarding dependency installation. Developers should validate that a recommended package exists and originates from a trusted source before installing it. \textbf{For researchers} and tool builders, our analysis points highlights a key gap in LLM tooling: there are no built-in mechanisms for verifying the validity or safety of recommended libraries. Future work should focus on mechanisms to incorporate runtime verification into LLM interfaces or development environments and flag unresolved or suspicious dependencies. Additionally, efforts to benchmark models for hallucination-driven security vulnerabilities remain underexplored yet critical. 

\section{Related Works}
\label{sec:related_works}
In this section, we discuss the related literature divided into two aspects. First, we discuss the works that have focused on dependency management and related challenges. Second, we discuss the works that report on the use of large language models as programming assistants.

\subsection{Dependency Management Challenges}
\label{sec:rec_systems}

While open source software libraries significantly reduce development time and costs~\cite{basili1996reuse, decan2019empirical, murphy2019predicts}, depending on numerous libraries introduces complexity and potential dependency management challenges~\cite{Latendresse_ASE2022, mujahid2021effective, wang2020empirical}. One such challenge is highlighted by Mujahid et al.~\cite{mujahid2023characteristics} who identify library selection as a critical aspect of dependency management. In their work, they surveyed developers from the npm ecosystem to qualitatively understand the characteristics of highly-selected libraries. Their results show that JavaScript developers believe that such libraries are well-documented, popular, and free of vulnerabilities. Building upon this work, our study leverages those same characteristics to categorize libraries used by various LLMs. Hauge et al.~\cite{hauge2009empirical} show that organizational library selection is an ad-hoc process that often relies on a combination of past experiences, expert advice, and online resources. This is further discussed in the work of Haenni et al.~\cite{haenni2013categorizing} where the authors surveyed developers about their decision-making when selecting a library to integrate into their application. Their findings show that, in general, developers do not apply a clear rationale when selecting libraries. Alternatively, developers opted for libraries that fulfilled the immediate task requirements. In this context, LLMs are increasingly being adopted to assist in programming tasks, but little is known about the characteristics of the libraries included in LLM-generated code. Our work empirically analyzes the types of libraries recommended by several state-of-the-art LLMs and assesses their popularity, maintenance, and licensing properties.

Dependency hell is a concept discussed in several studies~\cite{fan2020escaping, tanabe2018context, abdalkareem2020impact} and refers to a situation where a project has an excessive number of dependencies, and managing these dependencies becomes difficult and error-prone. Chen et al.~\cite{chen2021helping} discuss the impact of "trivial packages," referring to libraries implementing simple functionalities, on the npm ecosystem. Their survey highlights that developers struggle with the multiple dependencies introduced by these libraries, contributing to dependency hell. For instance, a developer reported the cascading effect of patching a deeply nested dependency, requiring updates throughout the dependency tree. Jafari et al.~\cite{jafari_update} investigate the relationship between npm library characteristics and the dependency update strategies adopted by its dependents. The authors report that the release status, the number of dependents, and the age of a library are the most important indicators of the dependency update strategy. Raemaekers et al.~\cite{raemaekers2011exploring} discuss the risks associated with the usage of third-party libraries. They identify key library attributes that could serve as risk indicators. Notably, they report that more popular libraries may be updated more frequently, which increases the chance for new bugs to get introduced into the codebase. These findings support our claim that LLMs could better support developers by providing critical information about the libraries they recommend, such as maintenance metrics (i.e., the number of dependencies, version update frequency, project age). Such transparency can help developers anticipate and manage potential maintenance challenges associated with the increased complexity of dependencies.

\subsection{LLMs in Software Development Workflows}
\label{sec:llms_dev_workflows}

LLMs are increasingly used by software engineering practitioners to perform various tasks, such as code generation, and have shown the potential to improve developer productivity~\cite{ross2023programmer, heitz2024evaluation, teubner2023welcome}. However, some studies have raised concerns about the reliability of LLM-generated code. For instance, Zhong et al.~\cite{zhong2024chatgpt} report common API misuse patterns found in popular LLMs. The study reveals that in the case of GPT-4, 62\% of the generated code contained API misuses. The authors argue that this is particularly problematic given that users of LLM code generation are generally not familiar with the APIs that LLMs generate code for, and cannot tell whether the provided code is correct. Similarly, a Vulcan Cyber article claims that ChatGPT hallucinated in almost 40\% of the programming questions it was asked~\cite{lanyado-2023}. 

Our findings diverge from those reported in the aforementioned studies. This discrepancy might be attributed to differences in experimental design. First, Zhong et al. employed "one-shot" or "few-shot" approaches, providing either irrelevant or relevant examples alongside the prompt. The Vulcan Cyber article describes a scenario where ChatGPT was instructed to generate code for a specific library and then repeatedly prompted for alternatives, potentially leading to hallucinated libraries. Second, we conducted our study on multiple more recent and capable LLMs and aimed to more closely simulate realistic developer-LLM interactions by issuing a single prompt per question. This approach aligns with findings from Jin et al.~\cite{jin2024can}, whose empirical study reveals that developers request code regeneration in only 3\% of conversations with ChatGPT. 

While LLMs are making significant advancements in software development, concerns regarding reliability and the potential for hallucinations remain. However, studies also highlight the potential of LLMs to improve developer productivity through functionalities like code completion and search. Ross et al.~\cite{ross2023programmer} developed an LLM-based programmer's assistant and evaluated their system on 42 participants. Their results reveal that participants were mostly positive about the assistant's potential for improving their productivity. Heitz et al.~\cite{heitz2024evaluation} evaluate and compare the performance of OpenAI's ChatGPT and Google's Gemini on programming code. They find that while the premium version of the models offers enhanced performance, their free counterparts remain highly relevant for a wide range of users in the context of software development. Also, the authors report that these models can significantly accelerate coding tasks and improve productivity, but necessitate a rigorous review, especially if the generated code is used in critical areas. Despite this, there is a gap in understanding how LLMs perform in the context of dependencies.

Our study directly addresses this gap by investigating the effectiveness of LLMs in recommending software libraries. We use real-world developer questions from Stack Overflow to analyze the libraries suggested by several recent LLMs (as of 2025) in a setting that reflects practical usage scenarios. This approach avoids the potential bias of controlled experimental prompts. By evaluating factors like library popularity and maintenance, licensing, and potential dependency challenges, our work aims to inform the development of more robust LLMs as programming assistants, and provide developers with recommendations and considerations when using such tools to streamline their workflows in the context of dependencies.

\section{Threats to Validity}
\label{sec:threats}

\noindent\textbf{Internal validity} considers the experimenter's bias and errors. A potential threat to the validity of our study is the manual classification of libraries and import failures. To mitigate this, all cases were reviewed collaboratively by the authors, and discrepancies were resolved through discussion. We did not employ multiple independent coders or inter-rater agreement measures (e.g., Cohen's kappa), as the classification relied on factual criteria (e.g., \texttt{cv2} is an alias for \texttt{opencv-python}), not subjective judgment. While this may still introduce some bias, we reduced the likelihood through validation by multiple authors. 

Moreover, the stochastic nature of LLMs means they can generate different outputs for the same input. To reduce this variability, we used a low temperature value (0.1) to produce near-deterministic completions. Finally, prompt formatting and parameter settings can affect LLM outputs. We used default parameters where applicable (except for temperature) and used the Stack Overflow questions for prompting as-is to mimic typical usage and ensure consistency across models.

\noindent\textbf{External validity} concerns the generalizability of our findings. One external threat relates to the time gap between the Stack Overflow questions and the current software ecosystem. Some libraries may have changed since the questions were posted. To control this, we used the latest available metadata (as of September 2025) when analyzing library characteristics. 

Another potential threat is data leakage, where questions from Stack Overflow may have been seen during model training. To assess this, we repeated our analysis on StackUnseen, a companion dataset to StackEval, designed to reduce overlap with LLM training data. A Mann–Whitney U test revealed no statistically significant differences in the characteristics of the libraries recommended, which suggests that data leakage did not affect our results. 

Finally, our findings may not generalize to all LLMs, programming languages, or datasets. The models evaluated are limited to a specific set of Python questions, and results may differ in other ecosystems, languages, or when using different prompts. 

\section{Conclusion}
\label{sec:conclusion}

In this paper, we assessed how effectively LLMs recommend and use software libraries during code generation. Our findings show that LLMs are generally robust in their ability to generate valid and appropriate library imports across a diverse set of Python programming questions. Models consistently recommend well-established third-party libraries that are popular, actively maintained, and permissively licensed. However, we find that this robustness is conditional. A non-trivial portion of recommended imports involved cases where the installable package name differed from the import name (e.g., \texttt{cv2} for \texttt{opencv-python}), or where module-level imports were used without referencing the parent library. In these situations, models often failed to provide installation guidance, which leaves developers to manually resolve dependencies. While technically correct, these omissions nonetheless negatively affect the usability of LLMs, especially for novice users or in automated workflows. 

From a practical standpoint, developers using LLMs for code generation should remain vigilant about dependency validation and license compliance. While most libraries were permissively licensed, we noted one case of a copyleft license, for which there is no way to know except for manual verification. Thus, LLM-generated code may accelerate software development, but still require post-processing to ensure correctness and legal compatibility. Tool builders and IDE designers could help close this gap by integrating dependency resolution checks, license-awareness feature, and automatic install command generation directly into LLM-assisted environments. 

Future work should explore techniques for making LLM outputs more transparent and execution-aware, such as prompting strategies that elicit installation instructions, or LLM pipelines augmented with static or runtime analysis. Further studies could also expand beyond the Python ecosystem to assess whether similar patterns hold across other ecosystems like JavaScript (npm), Java (Maven), or Rust (Cargo). Ultimately, improving the traceability, trustworthiness, and contextual completeness of LLM-generated code in the context of dependencies is critical for their safe integration into codebases.

\bibliographystyle{IEEEtran}
\bibliography{main.bib}

\clearpage
\appendix
\section{Appendix: Descriptive Statistics of Dependency Characteristics}

Table~\ref{tab:appendix_dependency_stats} presents the descriptive statistics (mean, standard deviation, median, minimum, and maximum) of the dependency characteristics aggregated across all models. These metrics complement the median-based comparisons shown in the main text (Table~\ref{tab:dependency_medians_all_and_per_model}) by providing a more detailed view of variability and outliers.

\begin{table}[h]
\centering
\caption{Descriptive statistics of dependency characteristics (aggregated across all models).}
\label{tab:appendix_dependency_stats}
\begin{tabular}{llrrrrr}
\toprule
\textbf{Category} & \textbf{Characteristic} & \textbf{Mean} & \textbf{Std. Dev} & \textbf{Median} & \textbf{Min} & \textbf{Max} \\
\midrule
\multirow{3}{*}{Popularity} 
& Forks        & 6,043.67  & 8,991.83  & 1,473   & 0   & 32,210 \\
& Stars        & 18,052.65 & 23,843.22 & 6,962   & 0   & 86,299 \\
& Dependents   & 10,092.03 & 19,435.10 & 1,140   & 0   & 77,540 \\
\midrule
\multirow{4}{*}{Maintenance} 
& Dependencies      & 10.49   & 19.50  & 1     & 0     & 85 \\
& SourceRank        & 20.76   & 8.00   & 23    & 3     & 31 \\
& Age (months)      & 99.52   & 50.74  & 95    & 3     & 130 \\
& Version Frequency & 2.41    & 4.21   & 0.64  & 0.01  & 16.32 \\
\bottomrule
\end{tabular}
\end{table}

\end{document}